# Mapping Citation Patterns of Book Chapters in the Book Citation Index

Daniel Torres-Salinas[a], Rosa Rodríguez-Sánchez[b], Nicolás Robinson-García[c]*, J. Fdez-Valdivia[b], J. A. García[b]

[a]EC3: Evaluación de la Ciencia y la Comunicación Científica, Centro de Investigación. Médica Aplicada, Universidad de Navarra, 31008, Pamplona, Navarra, Spain. Email address: torressalinas@gmail.com

[b]Departamento de Ciencias de la Computación e I. A., CITIC-UGR, Universidad de Granada, 18071 Granada, Spain. Email address: rosa@decsai.ugr.es; jfv@decsai.ugr.es; jags@decsai.ugr.es

[c]EC3: Evaluación de la Ciencia y la Comunicación Científica, Universidad de Granada, 18071 Granada, Spain. Email address: elrobin@ugr.es

**Abstract**

In this paper we provide the reader with a visual representation of relationships among the impact of book chapters indexed in the Book Citation Index using information gain values and published by different academic publishers in specific disciplines. The impact of book chapters can be characterized statistically by citations histograms. For instance, we can compute the probability of occurrence of book chapters with a number of citations in different intervals for each academic publisher. We predict the similarity between two citation histograms based on the amount of relative information between such characterizations. We observe that the citation patterns of book chapters follow a Lotkaian distribution. This paper describes the structure of the Book Citation Index using 'heliocentric clockwise maps' which allow the reader not only to determine the grade of similarity of a given academic publisher indexed in the Book Citation Index with a specific discipline according to their citation distribution, but also to easily observe the general structure of a discipline, identifying the publishers with higher impact and output.

## 1. Introduction

Books and book chapters - unlike articles and scientific journals - are document types that lack of sound and consolidated bibliometric measures. Nevertheless, only studies using small samples and focused in very specific fields can be found in the literature (for instance, Cronin, Snyder & Atkins, 1997; Lewison, 2001). The main reason and limitation for this is that there has not been an international and reliable multidisciplinary database with citation data. At the same time many approaches have been made using as a proxy their presence in libraries (White et al, 2009; Torres-Salinas & Moed, 2009; Linmans, 2010), book reviews (Zuccala & van Leeuwen, 2011) or other alternative databases as Google Books or Google Scholar (Kousha & Thelwall, 2009; Kousha, Thelwall & Rezaie, 2011), but none of them has been adopted unanimously by the bibliometric community. A possible reason for the lack of adoption of such measures may rely on the difficulty and time consuming efforts needed to obtain large data sets. Another characteristic of these studies is that they are usually focused on the Social Sciences and Humanities, as these fields rely heavily on monographs as their main communication channel (Hicks, 2004). This means a serious shortcoming to any bibliometric approach to these fields limited using only scientific articles and journals citation indexes (Archambault et al, 2006).

In this sense, the launch of the Book Citation Index by Thomson Reuters (hereafter BKCI) offers a new window of opportunities for the development of bibliometric indicators for these document types. Not only it is a new source for retrieving citation data, but it is part of the family of citation indexes developed by Thomson Reuters Web of Science, which are highly considered by the research community. This database was released in October 2010 and responded to a serious limitation acknowledged by Eugene Garfield himself, developer of the first citation indexes, who declared that the creation of the BKCI "would be an expected by-product of the new electronic media" (Garfield, 1996). However, due to its youth, few studies can be found in the literature referring to the use of the BKCI for evaluation purposes or describing its internal characteristics; coverage, limitations, etc. In fact, only two studies can be found. On the one hand, Leydesdorff & Felt (2012) analyze the citation rates of books, book chapters and edited volumes and compare the results offered by the BKCI with those of the other citation indexes. On the other hand, Torres-Salinas et al (2012) propose the development of a 'Book Publishers Citation Reports' and analyze the strengths and weaknesses of such attempt in the Social Sciences and Humanities fields. These types of seminal studies dissecting the coverage, caveats and limitations are considered of great regard as they serve to validate the accuracy and reliability of sources.

Meho & Yang (2007) consider that these studies can be divided into two groups: those related with the limitations of the Web of Science citation indexes and those that examine further sources. Although other databases have been used rather than the Web of Science for bibliometric purposes (Leydesdorff, Rotolo & Rafols, 2012), until 2004 no other database rather than this one provided bibliometric data (Bar-Ilan, 2008). Since then, with the launch of Scopus (Elsevier) and Google Scholar, many studies have emerged analyzing these alternative databases and their advantages and weaknesses when compared with the former (see e.g., Kulkarni et al, 2007; Moya-Anegón et al, 2007; or Kousha, Thelwall & Rezaie, 2011 for instance). However, a third group can be found which is related with the mapping and the analysis of the structure of the Web of Science citation indexes (Leydesdorff & Rafols, 2009). All of these types of approaches can be adopted when analyzing the BKCI. Nevertheless, this database allows a deeper analysis of books and book chapters than the ones available before. As long suggested by Line (1979), these may present a different behavior than that presented by journals. Now, this premise can be fully tested. Taking into account this background, the present study intends to unite the aforementioned perspectives. Firstly, we will analyze the citation phenomenon for the whole database. And secondly, we will employ science maps in order to deep on the information resources indexed by Thomson Reuters for the development of the BKCI.

Specifically, in this study we aim at analyzing the citation patterns of book chapters in the BKCI in four different disciplines: Humanities & Arts, Science, Social Sciences and Engineering &

Technology. Book chapters have been scarcely studied in the field of scientometrics, least with such a large data set as the one provided by the BKCI. Therefore it is interesting to analyze their citation behavior and characteristics as, although some studies have deepen on the citation patterns of books (Tang, 2008), none focus on book chapters. We take a novel approach using academic publishers as unit of analysis in order to perform a secondary analysis on the structure of the BKCI. A key issue when constructing citation indexes is the sources used or, as in this case, the selection of sources; which will determine the citation universe in which the index will be based. An interesting approach when evaluating books' impact is to focus on the prestige of their publishers, establishing an analogy with articles and journals (Giménez-Toledo & Román-Román, 2009; Torres-Salinas et al, 2012). In this paper we adopt such an analogy, applying theoretic information measures to map academic publishers according to their similarity with respect to the overall citation distribution of book chapters of the top 20 academic publishers in specific fields. We believe that this study offers a first approach to the BKCI database as the application of information theoretic measures allows us to identify the main publishers and their main characteristics by area, an important issue when studying and validating a new information resource. This methodology has already been successfully applied for benchmarking academic institutions (García et al, 2012).

Therefore, our aim is to develop what we have named 'heliocentric clockwise maps' as a means to describe the structure of the BKCI through book chapters' citation patterns. These maps allow the reader not only to determine the grade of similarity of a given academic publisher indexed in the BKCI with a specific discipline according to their citation distribution, but also to easily observe the general structure of a discipline, identifying the publishers with higher impact and output. They can even be used to detect deficiencies on the coverage of each field, offering a general overview of the strengths and limitations of the database. The paper is structured as follows. In Section 2 we describe the data recollection and processing. In Section 3 we describe the methodology employed giving the key points for understanding and interpreting the results. The results are shown in Section 4. Finally, in Section 5 we analyze thoroughly the results obtained, focusing on the behavior of academic publishers and we present our conclusions in Section 6. Also, in Appendix A we provide the reader with further information about the development of the information gain measure used for the construction of the heliocentric maps. Finally, we have included Complementary Material (available at http://hdl.handle.net/10481/22587) in order to enrich the analysis and provide the reader with further information.

**2. Data source and description of the database**

*2.1. Data source and processing*

In this study we map citation patterns of academic publishers with their book chapters. For this we selected the 2005-2011 study period. Records indexed as 'book chapters' according to the BKCI were downloaded in May 2012. The chosen time period is based on the availability of the data at the time of the retrieval, as then, the BKCI went back to 2005. Then, data was included into a relational database created for this purpose in order to process it and calculate the indicators. During data processing, publisher names were normalized as many had variants that differed as a function of the location of their head offices in each country. For instance,

Springer uses variants such as Springer-Verlag Wien, Springer-Verlag Tokyo, Springer Publishing Co, among others. In order to ease the analysis, the 249 subject categories to which records from BKCI are assigned, were also restructured into four disciplines. Aggregating subject categories is a classical perspective followed in many bibliometric studies when adopting a macro-level approach (Moed, 2005; Leydesdorff & Rafols, 2009). These aggregations are needed in order to provide the reader with an overview of the whole database. In this sense, we decided to cluster all subject categories into four macro areas (see tables 1-4, Complementary Material): Arts & Humanities, Science, Social Sciences and Engineering & Technology. This way we minimized possibilities of overlapping for records assigned to more than one subject category (12% of the total share was assigned to more than one area). Also, we consider that such areas are easily identifiable by the reader as they establish an analogy with the other Thomson Reuters' citation indexes (Science Citation Index, Social Science Citation Index and Arts & Humanities Citation Index). With the exception of Sciences, which due to the heterogeneity of such a broad area, was divided into two areas: Science and Engineering & Technology. In Table 1 we show the indicators calculated in this study in order to offer a general description of book chapters indexed in the BKCI.

**Table 1.** Set of indicators calculated and their definition for a general description of book chapters indexed in the Book Citation Index

| Indicator | Definition |
| --- | --- |
| Nr BC | Total number of book chapters for a given discipline |
| % BC from the Total Database | Percentage of book chapters of a given discipline considering the total share of the BKCI |
| Total Citations | Total number of citations received by all book chapters |
| Citations from the Total Database | Percentage of citations received by the book chapters of a given discipline considering the total share of the BKCI |
| Citation Average | Average of the number of citations received by book chapter |
| Citation Average Standard Deviation | Standard deviation of the average of the number of citations received by book chapter |
| Nr Academic Publishers | Total of academic publishers that contribute to the total share of book chapters of a given discipline |
| % BC – Top20 Publishers | Percentage of book chapters edited by the top 20 academic publishers considering the total share of a given discipline |
| Nr of Citations Most cited BC | Number of citations achieved by the most cited book chapter in a given discipline |
| % of Non-Cited BC | Percentage of book chapters which have remained uncited considering a given discipline |

*2.2. General description of the database*

The BKCI contains for the 2005-2011 period 367 616 book chapters (Table 2), which represent mainly the fields of Science and Social Sciences which cover 74% of the total share. The discipline less represented is Engineering & Technology (13% of the total share). However, one single discipline, Science, receives most of the citations (85%). For this field, book chapters receive an average of 3.32 citations each. The other three areas receive a total of citations of 9% for Social Sciences, 6% for Engineering & Technology and 3% for Arts & Humanities. If we focus on the presence of academic publishers by discipline, the BKCI includes 284 different publishers.

The BKCI shows an increasing distribution of book chapters per year for the study time period (Complementary material, table 5), as only 6.60% of the total share dates back to 2005, while 17.85% was published in 2011. When analyzing the distribution of the database according to the country of publication (Complementary material, table 12), we observe that 74.53% of the total share of book chapters indexed in the BKCI comes from only two countries, United States and England, showing a strong bias towards English speaking countries. In fact, the non-English speaking language with a greatest share of book chapters is Germany, reaching only 13.87% percent. Also France, Asiatic countries or Spanish speaking countries seem greatly underrepresented by the database. Regarding publishers' distribution (Complementary material, table 10): only three publishers gather 50.77% of the total database (Springer, 27.33%; Palgrave, 12.15%; Routledge, 11.29%).

**Table 2.** General indicators for book chapters in the Book Citation Index. 2005-2011

| INDICATORS | GENERAL DATABASE | Arts & Humanities | Science | Social Sciences | Engineering & Technology |
|---|---|---|---|---|---|
| Nr BC | 367616 | 95087 | 140444 | 130513 | 49316 |
| % BC From the total Database | 100% | 26% | 38% | 36% | 13% |
| Total Citations | 546510 | 16206 | 466405 | 49010 | 33645 |
| % Citations From the total Database | 100% | 3% | 85% | 9% | 6% |
| Citation Average | 1.49 | 0.17 | 3.32 | 0.37 | 0.68 |
| Citation Average Standard Deviation | 14.22 | 1.15 | 22.5 | 2.39 | 6.31 |
| Nr Academic Publishers | 284 | 127 | 191 | 134 | 80 |
| % BC – Top20 Publishers | 94% | 92% | 84% | 91% | 93% |
| Nr of citation most cited BC | 3359 | 159 | 3359 | 290 | 627 |
| % of Non-Cited BC | 83% | 92% | 74% | 87% | 85% |

In Table 2 we offer a general description of the contents of the BKCI and its distribution for book chapters and academic publishers among the four disciplines analyzed. Science is the area which includes more publishers (191), followed by Social Sciences, Arts & Humanities and finally, Engineering & Technology with 80 academic publishers. Despite these figures, only 20 publishers cover at least 84% of the total share (Science), being Engineering & Technology the discipline in which the largest 20 publishers cover the highest percentage of the total share (93%). Another important issue worth mentioning is the high rates of uncitedness. 92% of the book chapters belonging to Arts & Humanities remained uncited, followed by Social Sciences (87%) and Engineering & Technology (85%). Science is the discipline with the lowest rate of uncitedness with 74% of book chapters uncited.

**3. Methodology for mapping academic publishers: Information Gain**

One of the goals of this paper is to provide a visual representation of the relationship among citation patterns of book chapters published by top academic publishers in four disciplines. To this aim, two different problems have to be solved: Firstly, we need a reasonable method to characterize the contribution of book chapters which were published by certain academic publisher; and secondly, we have to be able to measure what is the amount of relative information between such characterizations. In this section we give the key points for understanding and interpreting the application of the information gain methodology in this study. For a more exhaustive presentation of this methodology the reader is referred to Appendix A and to García et al (2012) where this methodology is applied in a bibliometric context for benchmarking academic institutions.

Information gain or Kullback-Leibler divergence (Kullback & Leibler, 1951) is a measure that allows us to select the academic publishers that contribute with more information to a given discipline. It compares two distributions; a true probability distribution $p(x)$ and an arbitrary probability distribution $q(x)$, and indicates the difference between the probability of $X$ if $q(x)$ is followed, and the probability of $X$ if $p(x)$ is followed. Although it is sometimes used as a distance metric, information gain is not a true metric since it is not symmetric and does not satisfy the triangle inequality (making it a semi-quasimetric). In this paper, the true probability distribution $p(x)$ is represented by the citation distribution of disciplines, to which we refer as standard disciplines, while the arbitrary probability distribution $q(x)$ is represented by the citation distribution of academic publishers.

If we predict the similarity between the standard discipline and academic publishers based on their information gain, then the minimum value of information gain between an academic publisher and the standard discipline leads to the most alike publishers to the citation distribution of the discipline. The objective is twofold: firstly, to characterize the information gain between two probability distributions (representing each one of the academic publishers as well as the standard disciplines) with a minimal number of properties which are natural and thus desirable; and secondly, to determine the form of all error functions satisfying these properties which we have stated to be desirable for predicting discipline-publisher dissimilarity. This analysis allows identifying similar and dissimilar distribution from a given one, but it does not explain the reasons for such dissimilarity. It is based on a formal approach for predicting visual target distinctness in Computer Vision (García et al, 2001)". These probability distributions are represented through citation histograms.

**Figure 1.** Description of the development of the Heliocentric Clockwise Maps

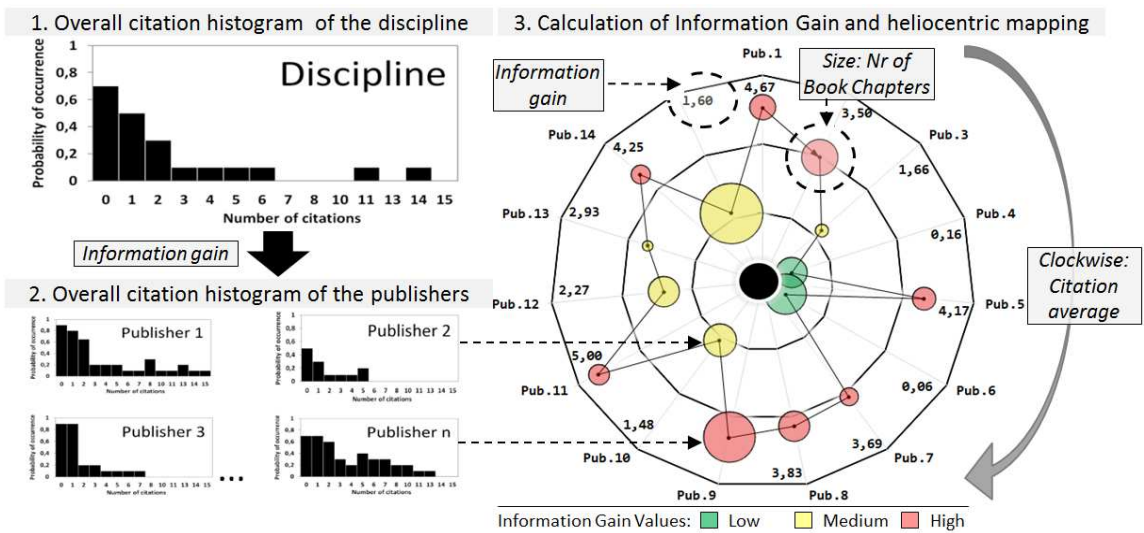

The citation pattern of publishers in specific disciplines can be characterized statistically by citations histograms. To this aim, for each publisher, we can compute the probability of occurrence of book chapters receiving a number of citations in different intervals. Although citations histograms do seem a good solution for visualizing the information gain between distributions, our aim is to offer a global picture of the whole discipline. Therefore, we developed what we have called the 'Heliocentric Clockwise Maps' (Figure 1). These maps are interpreted as follows. The center of the circle would be the distribution to which the other distributions are compared; in our case it would represent the standard discipline's distribution. The dots surrounding the centre of the circle would represent the publishers' distributions $q(x)$. Therefore, the ones closer to the center (lower information gain values) would show a more similar pattern to that of the discipline and the ones further way (higher information gain values) would perform more differently. The size of the dots represents the number of book chapters of academic publishers. The maps are named clockwise because the order of the publishers represents their citation average. Therefore, the publisher at the top of the circle has the highest citation average and so on, until the one on its left side which shows the lowest citation rate. This allows the reader to better interpret the meaning of more or lesser information gain (higher citation rates or lower citation rates) and the relation between different indicators. Only top 20 publishers were considered in the construction of the heliocentric clockwise maps. This decision is based on the fact that the top 20 publishers of each discipline cover more than 80% of the total output in all cases.

### 4. Results

*4.1. Histograms and calculation of information gain*

In Figure 2 we show the citation distribution histograms by discipline. These histograms represent the citation probability distribution of book chapters. In all cases we see that such distribution follows the same pattern. Zero has the highest probability. This distribution is more pronounced in Arts & Humanities (0.92) (Figure 2.A), and less pronounced in Science (0.721) (Figure 2.B). Except in the case of Science, there is practically no probability of a chapter receiving more than 3 citations.

**Figure 2**. Histogram representing probability of citations received by Book Chapters in the Book Citation Index in four different disciplines. Period 2005-2011.

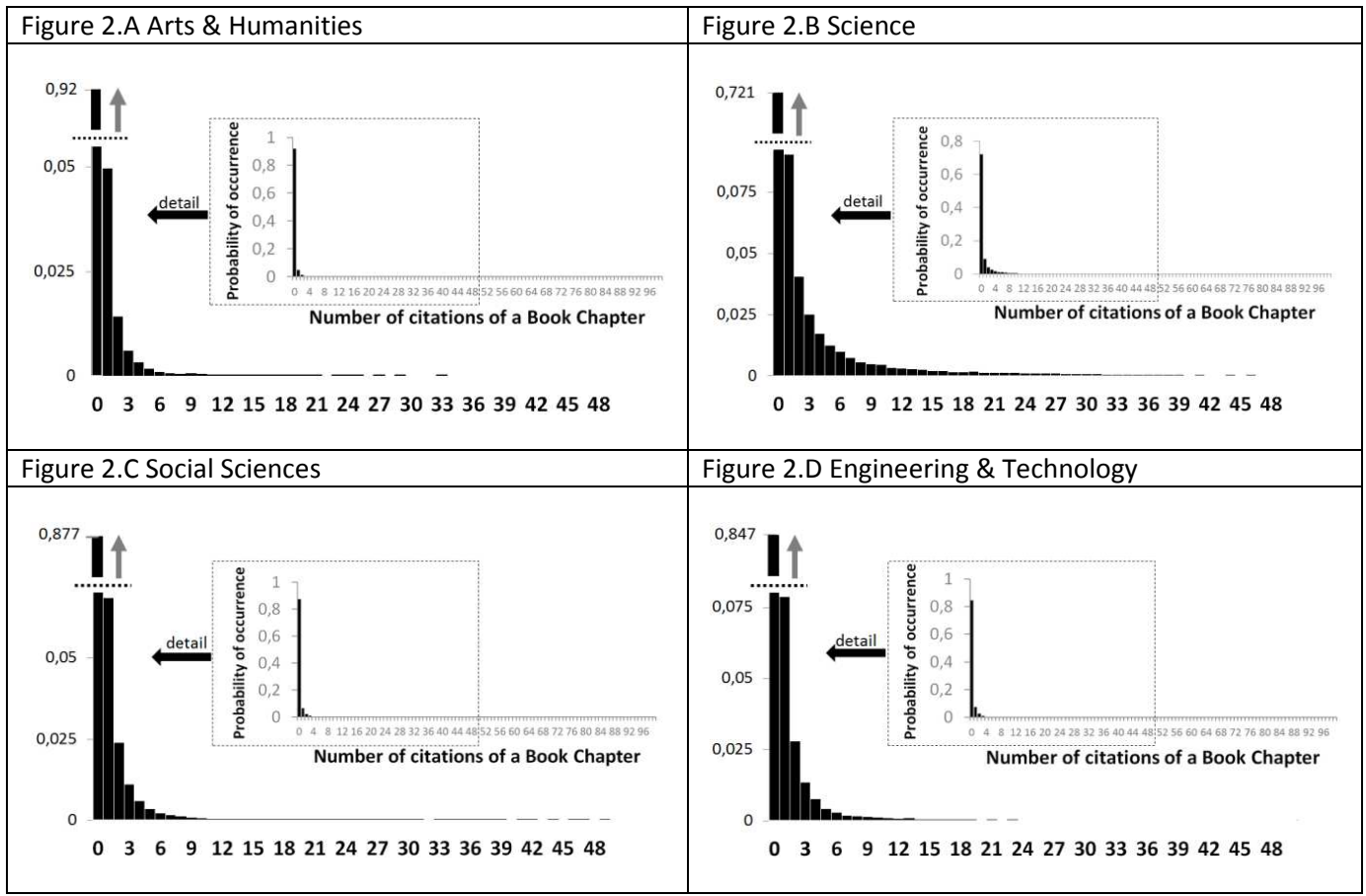

Another interesting observation is that in all cases, citation patterns (with the number of citations greater than 0) clearly follow a Lotkaian distribution. That is, they follow a function such as:

$$\varphi(n) = \frac{C}{n^\alpha}$$

(1)

where $n$ is the number of citations, with $n > 0$, and $\alpha$ a positive constant equal or higher than 1. In our application, $C$ and $\alpha$ are constants depending on the specific discipline (Complementary Material, Table 11). Here, Lotka's law states that 'the number (of book chapters) receiving $n$ citations is about $1/n^\alpha$ of those receiving one; and the proportion of all book chapters that receive a single citation, is about ($C$ times 100) percent". This means that out of all the book chapters in a given discipline, ($C$ times 100) percent will have just one citation, and ($C/2^\alpha$ times 100) percent will have two citations. ($C/3^\alpha$ times 100) percent of book chapters will have three citations, and so on. Lotka's Law, when applied to one discipline over a fairly long period of time, can be accurate in general, but not statistically exact (Complementary Material, Table 5).

Next, we show the histograms of publishers with the maximum gain and the minimum gain values for each discipline (Figure 3). Contrarily to what occurred before, the histograms of publishers with maximum gain do not always follow a Lotkaian distribution. If we approximate Lotka's Law in the case of maximum information gain the error will be much higher. A minimum gain means a greater similarity to the standard discipline and a maximum gain a lesser similarity. This must not be interpreted as having a higher or lower citation average. In fact, not always the academic publisher with a minimum information gain has a higher citation average than the one with maximum information gain. This occurs in the case of Engineering & Technology as well as in the case of Science, where the publisher with a maximum gain (Annual Reviews in both cases), shows higher citation rates than the one that performs more similarly to the discipline, Springer (which shows minimum gain values).

**Figure 3**. Histograms representing probability of citations received by publishers in the Book Citation Index in four different disciplines

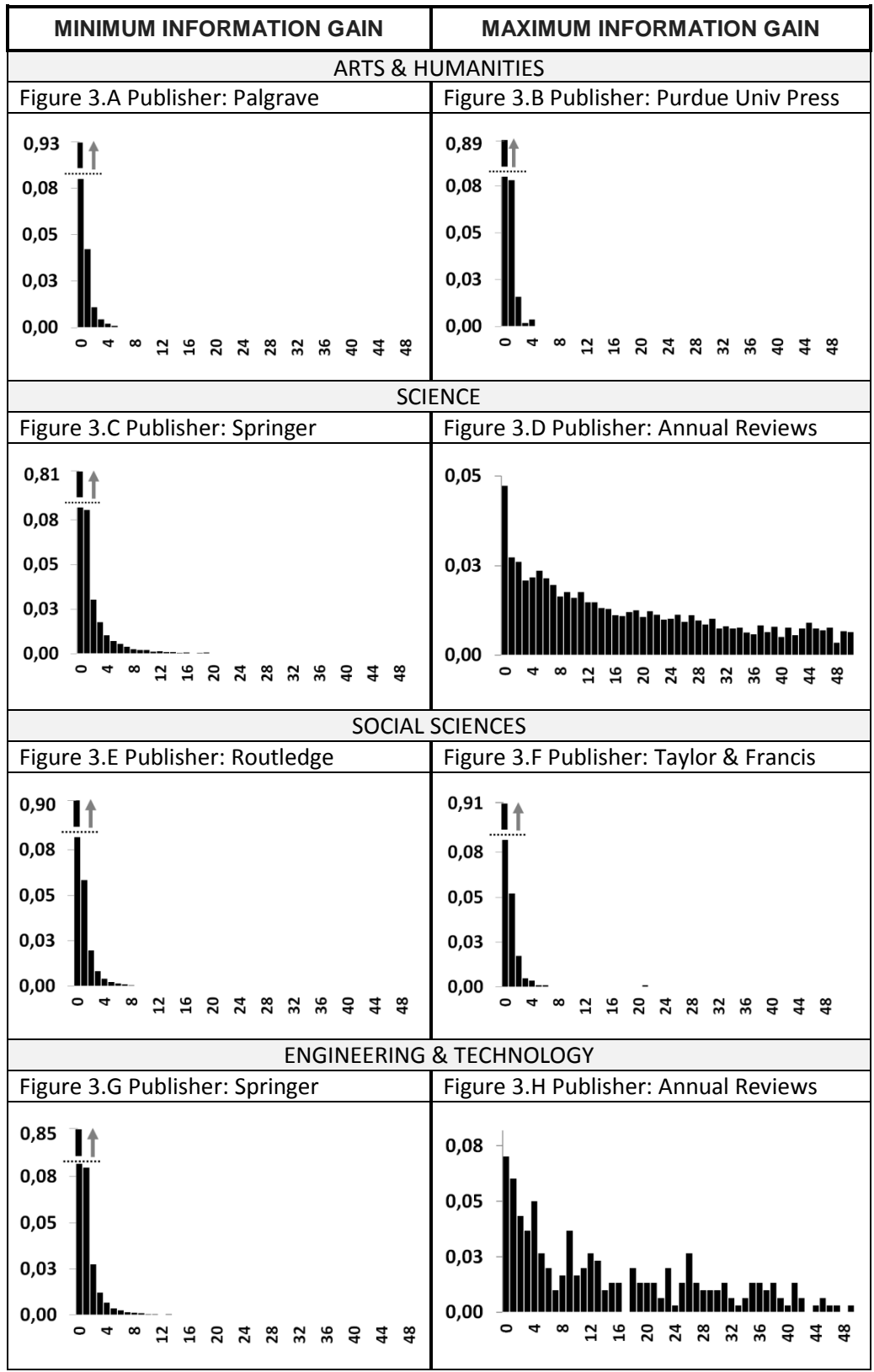

In Arts & Humanities we observe that Palgrave (Figure 3.A) is the publisher that follows a more alike distribution to that of the discipline (Figure 2.A), while Purdue (Figure 3.B) shows a different distribution. The same occurs in Social Sciences where Routledge (Figure 3.E) is the most similar publisher to the discipline while Taylor & Francis (Figure 3.F) is the least similar publisher. The aforementioned case of Annual Reviews is the one which shows a more anomalous behavior for both disciplines; Science and Engineering & Technology (Figures 3.D, 3.H), as its histogram shows a completely different distribution to that of the discipline (Figures 2.B, 2.D). In fact, while there is practically no probability of being cited more than 12 times in the discipline of Engineering & Technology, the distribution of Annual Reviews indicates that book chapters belonging to this publisher have reasonable probabilities of being cited even 48 times. This pattern prevents us from considering its books and book chapters as such but as review articles. This belief is reinforced when analyzing its records as they are indexed as book chapters and articles and do not include an ISBN but an ISSN. This evidence made us remove this publisher from our study based on Heliocentric Clockwise maps. However, in section 5 of Complementary Material we show the figures for each area in which Annual Reviews would have been included if it had not been excluded.

*4.2. Comparing publishers Information Gain using Heliocentric Clockwise Maps*

Figures 4-7 show the Heliocentric Clockwise Maps of each discipline representing the largest academic publishers. The data under these figures is available in tables 6 to 9 in the Complementary Material. These are ordered clockwise attending to their citation average; the publisher at the top of the map is the one with the highest citation average and so on. Therefore, the academic publisher on its left side is the one with the lowest citation average. For instance, in the case of Arts & Humanities (Figure 4), MIT Press is the publisher with the highest citation average (0.46), while EJ Brill is the one with the lowest average (0.02). Colors represent the grade of information gain publishers have according to the standard discipline.

Generally, we observe that the publisher with the highest citation average usually has a high information gain value and has a small size, as it occurs with MIT Press in Arts & Humanities (Figure 4) as well as in Social Sciences (Figure 6). This also happens for Elsevier in Engineering & Technology (Figure 7). However, it does not occur in Science where the two publishers with higher citation averages (Elsevier and Cambridge University Press) have intermediate information gain values. It is the third publisher with the highest citation average, The Geological Society of America Inc, the one with the highest information gain.

**Figure 4**. Heliocentric Clockwise Map representing the Information Gain for top academic publishers in Arts & Humanities in the Book Citation Index. Period 2005-2011.

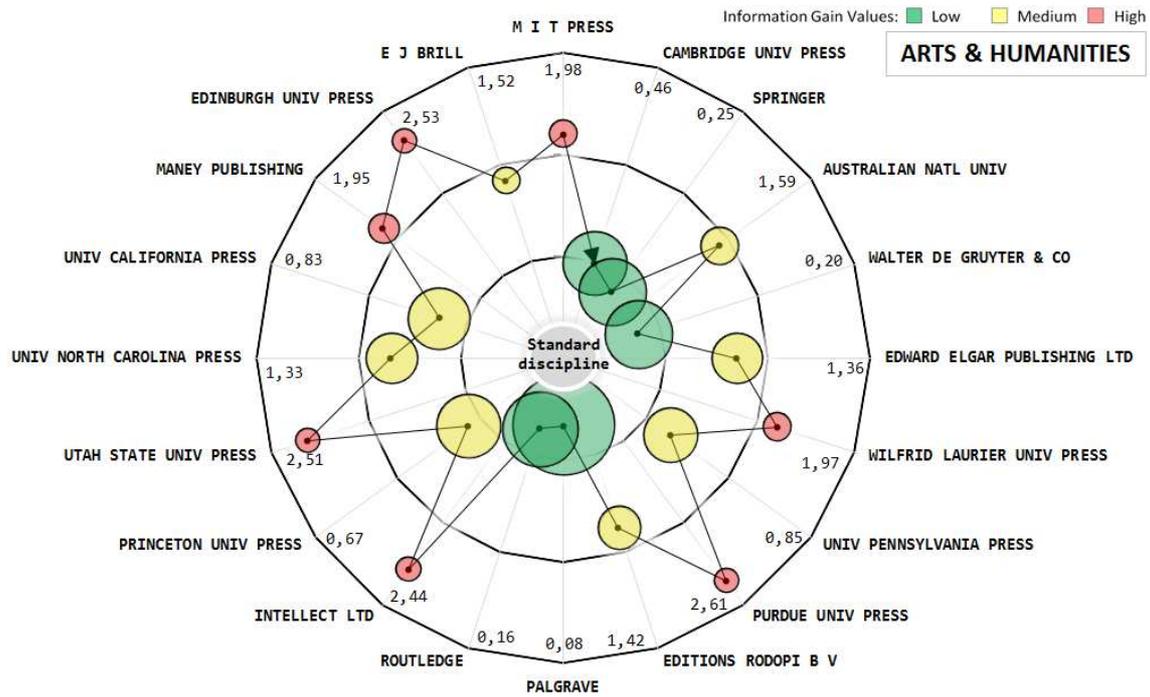

**Note**: Citation average values ranged from 0, 46 (MIT PRESS) to 0, 02 (EJ BRILL) | Volume values ranged from 22 444 (PALGRAVE) to 497 (PURDUE UNIVERSITY PRESS). Colors representing the Information Gain Values are introduced to aid the reader on the interpretation of the map.

In three disciplines, Engineering & Technology, Social Sciences and Arts & Humanities, it can be observed that the smaller academic publishers, in terms of research output are also those with a higher information gain and therefore, less alike with the discipline. On the other side, the biggest publisher shows lower information gain values and follows a more similar pattern to that of the standard discipline.

**Figure 5**. Heliocentric Clockwise Map representing the Information Gain for top academic publishers in Science in the Book Citation Index. Period 2005-2011.

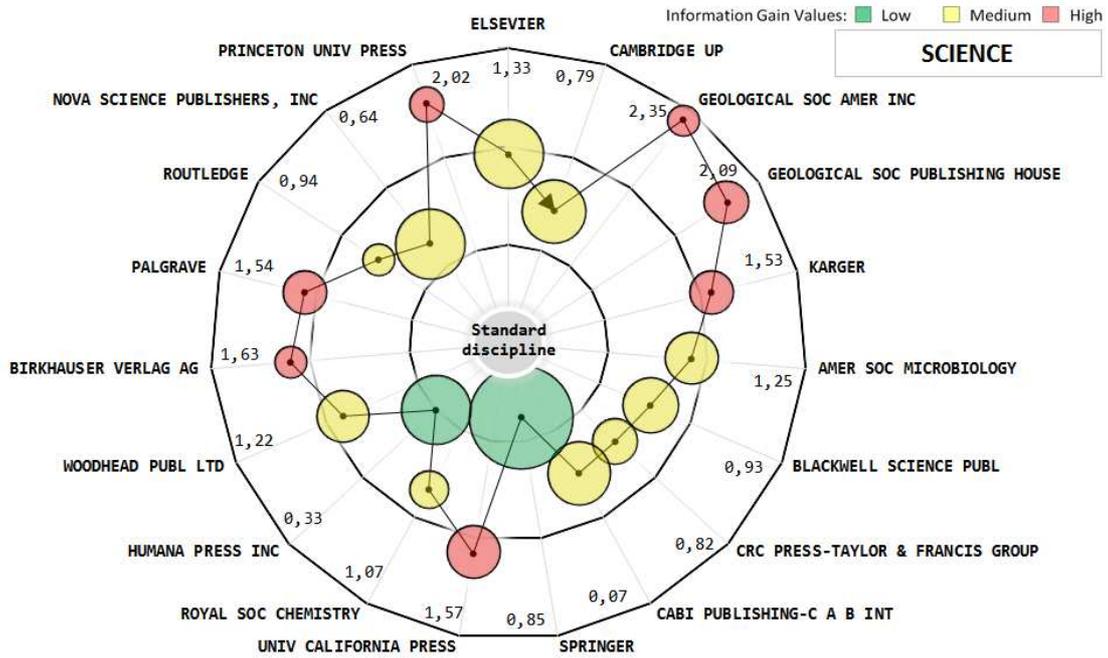

**Note**: Citation average values ranged from 9.07 (ELSEVIER) to 0.06 (PRINCETON UNIVERSITY PRESS) | Volume values ranged from 54 542 (SPRINGER) to 1 197 (BIRKHAUSER VERLAG AG). Colors representing the Information Gain Values are introduced to aid the reader on the interpretation of the map.

However, this behavior is not observed in the case of Science (Figure 5). There seems to be no such relation between size and information gain. In fact, we observe that publishers are more homogeneously distributed, with more similar citation patterns to that of the standard discipline. The behavior of the smaller academic publishers in terms of their book chapters' citation probability is different in this discipline to that of the other three (Engineering & Technology, Social Sciences and Arts & Humanities).

**Figure 6**. Heliocentric Clockwise Map representing the Information Gain for top academic publishers in Social Science in the Book Citation Index. Period 2005-2011.

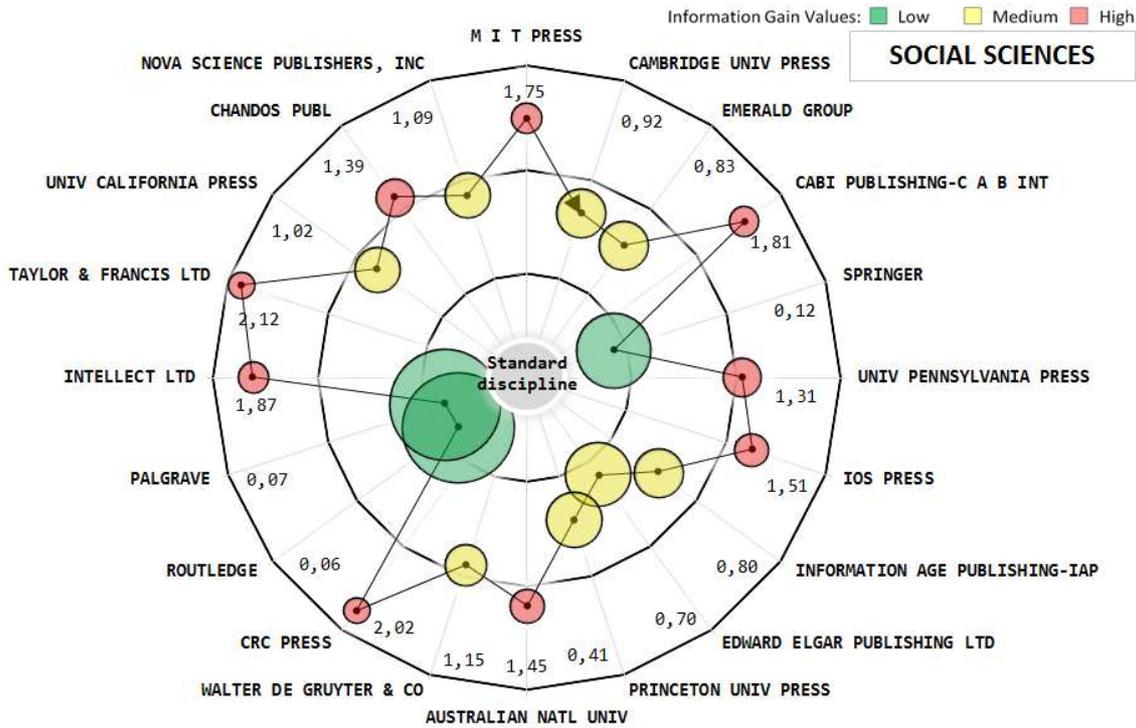

**Note**: Citation average values ranged from 0.85 (MIT PRESS) to 0.07 (NOVA SCIENCE) | Volume values ranged from 28 849 (ROUTLEDGE) to 800 (TAYLOR & FRANCIS). Colors representing the Information Gain Values are introduced to aid the reader on the interpretation of the map.

Regarding the presence of academic publishers in each discipline, we observe that the discipline of Engineering & Technology (Figure 7) is greatly unbalanced. Springer dominates the area accumulating approximately 62% of the total share, that is; 28 000 book chapters of the total of 40 000 belong to this publisher. Other disciplines may also be unbalanced but not to such extent. This fact influences the distribution of citation probability for book chapters in this discipline. If we compare the histogram of the discipline (Figure 2) and the one of Springer (Figure 3), we observe that is practically identical. This publisher has the minimum information gain value with 0.01, which means that its citation pattern is almost equal to the one of the standard discipline.

**Figure 7**. Heliocentric Clockwise Map representing the Information Gain for top academic publishers Engineering & Technology in the Book Citation Index. Period 2005-2011.

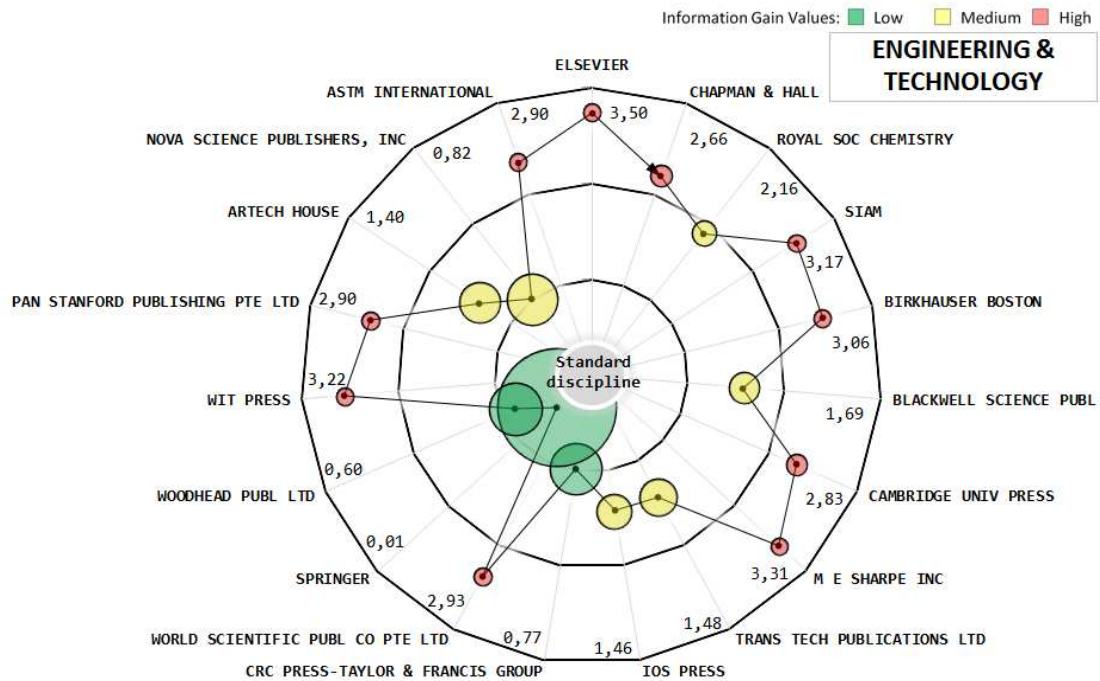

**Note**: Citation average values ranged from 4.76 (ELSEVIER) to 0.05 (ASTM INTERNACIONAL) | Volume values ranged from 28 471 (SPRINGER) to 236 (WIT PRESS). Colors representing the Information Gain Values are introduced to aid the reader on the interpretation of the map.

## 5. Discussion

In this paper we have applied the information gain measure to citations distributions of book chapters in the BKCI in order to analyze their citation patterns. For this, we divided the total output of the BKCI in four disciplines which are Science, Engineering & Technology, Arts & Humanities and Social Sciences. Then, we calculated the citation probability distribution of each academic publisher in the BKCI and the citation probability distribution of each of these four fields. This way, the information gain measure was calculated as for the top 20 most productive publishers of each discipline as they cover at least 84% of the total share of each discipline. Finally, we constructed the so-called 'Heliocentric Clockwise Maps' which visualize a discipline's structure allowing the reader to easily analyze the main academic publishers of a discipline, the ones with more impact, flaws on the BKCI coverage or the relation between specialization in a certain field and impact.

When analyzing the pattern of book chapter citations we observe that in all disciplines the distribution is highly skewed. In fact, the distributions are very similar to those described by Seglen (1999). Also, different fields show different citation behaviors. While the skewness and the uncitedness rate are higher for Arts & Humanities, in Science they are lower, following a similar phenomenon to that described by Hamilton (1991). Therefore, an evident conclusion would be that the citation distributions of book chapters follow a standard pattern, similar to the one followed by scientific publications. This statement is also corroborated by the fit of Lotka's law to the citation histogram of each specific discipline (Figure 8).

**Figure 8**. Fitting Lotka's law to histogram representing probability of citations received by Book Chapters in the Book Citation Index in four different disciplines. Period 2005-2011.

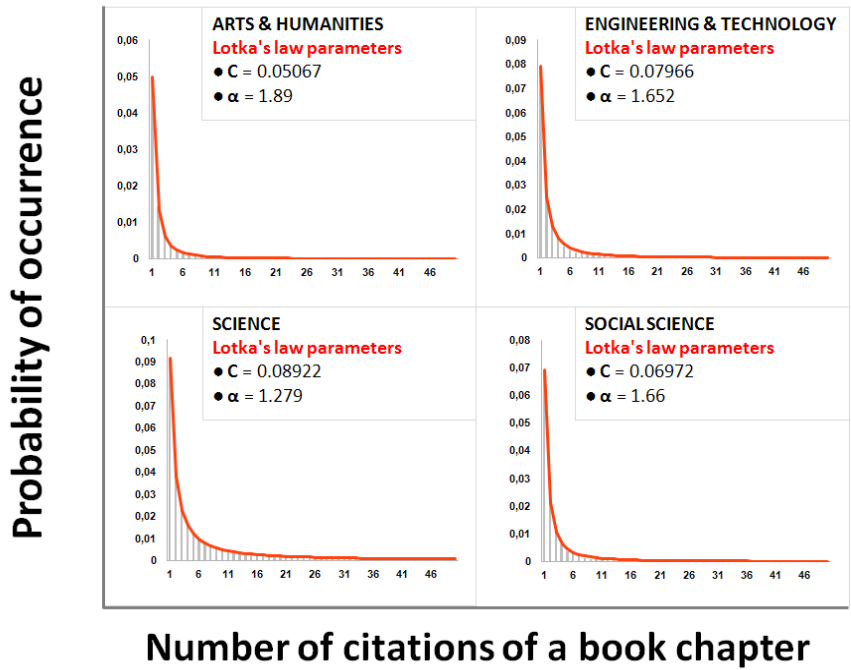

Also three characteristics seem to be highly related; citation average, information gain and publisher output. That is, those academic publishers with higher citation average also have a high information gain value and are usually small publishers. This happens in all disciplines except in Science, where the size of the publishers is more balanced and does not seem to relate with any other of the other two characteristics. Elsevier or Cambridge University Press specially but also other publishers such as California University Press or American Society for Microbiology show a considerable output but still have high information gain values. This behavior leads us to the conclusion that although output and information gain are indeed related, the heliocentric maps still allow us to spot outliers despite of their volume. In this sense, we believe that the influence of size may happen only when areas are not well balanced according to the publishers' output. In Social Sciences, Arts & Humanities and Engineering & Technology, those publishers with higher citation average and information gain are usually smaller than the rest of publishers. This phenomenon may be due to the influence the larger academic publishers have on the standard discipline, distorting its citation distribution. In fact, we observe in Engineering & Technology the great influence of Springer which has an almost identical pattern than the one of the standard discipline.

However, an exception has been noted: Annual Reviews. This publisher has an anomalous behavior in terms of information gain compared with the rest of the publishers. As suggested by Torres-Salinas et al (2012), it may well be because of its nature, more similar to that of journals than monographs. In fact, when removing this publisher from our analysis and we identify the publisher with the highest citation average and informatics gain, we observe that for Engineering & Technology, Elsevier stands up. Also, it verifies the third characteristic mentioned above, which is that it has a small size when compared with the output of the rest of publishers in the discipline. Also, another issue raises as Elsevier excels in two disciplines (Science and Engineering & Technology).

Another interesting issue is the behavior followed by Springer, Palgrave, Routledge and Nova Science in all disciplines. These publishers have a big size in all cases (in terms of output), perform with low information gain and medium-low citation average values for all disciplines. However, Springer stands out of the three regarding its citation average varies depending on the discipline: it is relatively high in Arts & Humanities, but it performs with low values in Engineering & Technology. While the other publishers citation average always shows medium-low values.

## 6. Conclusions

In this paper we have introduced a representation improvement for analyzing the citation patterns of book chapters in the BKCI in four different disciplines: the 'Heliocentric Clockwise Maps'. These maps represent publishers' alikeness to a standard discipline according to the information gain values between the book chapters' citation probability distribution of academic publishers and the overall distribution of the discipline. We have analyzed the BKCI according to the academic publishers in Science, Engineering & Technology, Social Sciences and Arts & Humanities. The citation distribution of book chapters follows the same pattern than the one in journals, behaving as suggested in Lotka's law and demonstrated in the case of the latter by Egghe (2005), with highly skewed distributions (Seglen, 1999). In this sense, further analyses in this line of work as they would deepen on these similarities between books, book chapters and articles' citation behavior. Normally, publishers with high citation average are the ones less alike of the discipline and have a relatively small size. Annual Reviews presents an outlier pattern in this sense which could be attributed to a behavior more closely linked to that of journals rather than monographs, as suggested elsewhere (Torres-Salinas et al, 2012), warning against its use when analyzing the BKCI.

Following this line of thought, we observe that the largest publishers across all fields are Springer, Routledge, Palgrave and Nova Science. Whilst they do not perform very well regarding their citation average, they influence greatly the citation pattern of all disciplines. In fact, Springer shows low information gain values in all disciplines. For instance, in Engineering & Technology Springer is not only the largest publisher, but its information gain value tends to zero, concluding that this discipline is poorly covered by the BKCI as it is represented by few or even just one publisher. This leads to the conclusion that, unlike in journals citation indexes, a large publisher may well distort the final picture of the BKCI, therefore in order to obtain a balanced coverage of a discipline, a balanced coverage of publishers is also needed. These maps may be used not just for analyzing the citation pattern of book chapters and academic publishers but also as a methodology for studying the coverage of the BKCI. Finally, we believe that the present study will contribute to the understanding of the BKCI and its limitations for future bibliometrics analyses; offering not only an overview of its coverage but also underlining its flaws.

**Acknowledgments**

Thanks are due to the reviewers for their constructive suggestions. This research was sponsored by the Spanish Board for Science and Technology (MICINN) under grant TIN2010-15157 co financed with European FEDER funds. Nicolas Robinson-García is currently supported by a FPU grant from the Spanish Ministerio de Economía y Competitividad.

**Complementary Material**

A document containing further information on the analyses conducted in this study has been elaborated and is freely available at the following link: http://hdl.handle.net/10481/22587.

**Appendix A. Basic axiomatic characterization of a measure of information gain**

It often happens that the contribution of book chapters published by certain academic publisher in specific disciplines cannot be accurately determined due to various reasons: some of the details may not be observable or the researcher who makes an attempt to investigate the impact may not take all the relevant factors governing the contribution of book chapters into consideration. Under such circumstances, the impact of book chapters published by some academic publisher can be characterized statistically by histograms of number of citations. For instance we can compute the probability of occurrence of book chapters with a number of citations in the interval $[l_i, l_i + \Delta l[$, with $i = 0, 1, \cdots, n$, for each academic publisher, and where $l_i, l_i + \Delta l$.

Let us assume the discrete probabilities associated with a reference publisher $R$ and another of input $I$ as those given by $P$ and $Q$, respectively, but what is the amount of relative information between $P$ and $Q$? To answer this kind of questions, a large number of measures have been developed by Jeffreys (1946), Kullback & Leibler (1951), Renyi (1961), and others. This makes it very difficult when choosing the criteria in order to see which one suits better. In order to do so, it is important to know which postulates and properties should be satisfied by the information theoretic measure.

Here we present a basic axiomatic characterization of a measure of information gain between an input academic publisher $I$ and another of reference $R$, where information gain measures the degree of dissimilarity between these two academic publishers. If we predict the similarity between academic publishers based on their information gain, then the minimum value of information gain between two publishers leads to the most similar ones. The objective is twofold: firstly, to characterize the information gain between two probability distributions (representing each one of the academic publishers) with a minimal number of properties which are natural and thus desirable; and secondly, to determine the form of all error functions satisfying these properties which we have stated to be desirable for predicting publisher-publisher dissimilarity.

The first postulate states a property of how unexpected a single event of an academic publisher was.

**Axiom 1.** *A measure $U$ of how unexpected the single event "a book chapter with a number of citations in the interval $[l_i, l_i + \Delta l[$ occurs" was, depends only upon its probability $p$.*

This means that there exists a function $h$ defined in [0, 1] such that $U$ ("a book chapter with a number of citations in the range $[l_i, l_i + \Delta l[$ occurs") $= h(p)$. This is a natural property because we assume that the academic publishers are characterized by discrete probability distributions (e.g., histograms of number of citations).

Our second postulate is formulated to obtain a reasonable estimate of how unexpected an academic publisher was from some probability distribution by means of the mathematical expectation of how unexpected its single events were from this distribution.

**Axiom 2.** *An estimate of how unexpected the impact of book chapters published by a reference academic publisher was from certain probability distribution is simply defined as the mathematical expectation of how unexpected its single events "a book chapter with a number of citations in interval $[l_i, l_i + \Delta l[$ occurs" were from that distribution.*

The following postulate relates the estimate of how unexpected the reference academic publisher was from an "estimated" distribution and the estimate from the "true" distribution.

Let $p(l/R)$ and $p(l/I)$ be the probability of occurrence of a publication with a number of citations in the interval $[l_i, l_i + \Delta l[$ for a reference publisher R and the input one I, respectively. Suppose that every possible observation from $p(l/R)$ is also a possible observation from $p(l/I)$.

If the single events of the reference publisher R are characterized by an "estimated" distribution $Q = \{p(l_i/I) | i = 0, 1, \cdots, n\}$, then the function $h(p(l_i/I))$, with $i = 0, 1, \cdots, n$, returns a measure of how unexpected each single event "a publication with a number of citations in the interval $[l_i, l_i + \Delta l[$ occurs" was from $Q$. Thus, assuming that $P = \{p(l_i/R) | i = 0, 1, \cdots, n\}$ is the "true" probability distribution of the reference academic publisher $R$, we have that:

**Axiom 3.** *The reference academic publisher $R$ with "true" probability distribution $P$ is more unexpected from an "estimated" distribution $Q$ than from the "true" distribution $P$.*

The following inequality expresses how the reference academic publisher is more unexpected when it is characterized by $Q$ than when is characterized by $P$:

$$U_P((Q)) \geq U_P(P)$$

(A.1)

with $U_P(Q)$ and $U_P(P)$ being estimates of how unexpected the reference academic publisher was from the "estimated" distribution $Q$ and from the "true" distribution $P$, respectively.

The true distribution $Q$ of the input academic publisher $I$ may be interpreted as an estimated distribution of the reference publisher $R$ (with "true" distribution $P$). Thus, we can define a measure of information gain of the reference publisher from the input one by the difference between the estimate of how unexpected the reference publisher was from $Q$ and that from $P$.

**Definition 1: A measure of information gain between academic publishers.** Given the reference academic publisher $R$ with "true" probability distribution $P = \{p(l/R)\}$, a measure of the information gain of the reference publisher $R$ from the input one I with "true" distribution $Q = \{p(l/I)\}$, is:

$$\varepsilon(P,Q) = U_P(Q) - U_P(P)$$

(A.2)

With $U_P(Q)$ and $U_P(P)$ being estimates of how unexpected the reference academic publisher was from $Q$ and $P$, respectively. $U_P(Q)$ and $U_P(P)$ are defined as given in Axiom 2, and such that satisfy the inequality (A.1) in Axiom 3.

The following result serves to determine the form of the measure $\varepsilon(P,Q)$.

**Theorem 1.** Let $\varepsilon(P,Q)$ be a measure of information gain for the discrimination between two academic publishers as given in Definition 1, i.e.,

$$\varepsilon(P,Q) = U_P(Q) - U_P(P)$$

(A.3)

with $P = \{p(l/R)\}$ and $Q = \{p(l/I)\}$. Then, the measure of relative information $\varepsilon$ is equal to the Kullback-Leibler's information function (Kullback, 1978) between $P$ and $Q$ up to a nonnegative multiplicative constant, i.e.,

$$E(P,Q) = aE_p\left[\log\frac{P}{Q}\right]$$

(A.4)

with $a \geq 0$ and $E_p$ denoting the mathematical expectation.

Proof. See Theorem 1 in (Garcia et al., 2001)

In conclusion, any measure of information gain between two academic publishers that satisfies Axioms 1, 2, and 3 has to be of the form of the Kullback-Leibler information function up to a nonnegative multiplicative constant.

Following the same approach, the information gain given in Definition 1 can also be used to measure the relative information between the overall citation histogram of the discipline (e.g., Science) and the citation histogram of certain publisher. In this case, the information gain measures the dissimilarity between discipline and publisher.